\documentclass[conference]{IEEEtran}
\usepackage{amsmath,amsfonts}
\usepackage{algorithmic}
\usepackage{algorithm}
\usepackage{array}
\usepackage[bookmarks=false,hidelinks]{hyperref}
\usepackage{textcomp}
\usepackage{stfloats}
\usepackage{url}
\usepackage{verbatim}
\usepackage{graphicx}
\usepackage{cite}
\usepackage{booktabs}
\usepackage{multirow}
\usepackage{hyperref}
\usepackage{listings}
\usepackage{booktabs}
\usepackage{multirow}
\usepackage{tabularx}
\usepackage{array}
\usepackage{subcaption}
\usepackage{graphicx}
\usepackage[T1]{fontenc} 

\begin{document}

\title{Uncertainty Signals for Network Intent Translation: \\Risk Ranking and Ambiguity Localization}
\author{
\IEEEauthorblockN{Ala' A. Alsamarneh$^{1}$, and Omar Alhussein$^{1}$}
\IEEEauthorblockA{
$^{1}$Department of Computer Science, College of Computing and Mathematical Sciences \\
Khalifa University, Abu Dhabi, United Arab Emirates\\
100067691@ku.ac.ae, omar.alhussein@ku.ac.ae
}
}

\maketitle

\begin{abstract}
Intent-based networking realization starts by translating high-level intents into low-level network configurations. Recent approaches have shifted toward LLM-based translation. Despite promising results, most studies focus on translation accuracy and overlook risks associated with deploying the resulting configurations. In this work, we investigate the pre-deployment translation risk of LLM-generated configurations by analyzing the model’s uncertainty. We propose to use two uncertainty signals, namely sampling-based predictive uncertainty for translation-risk ranking and token-level entropy for ambiguity-source localization. We evaluate these signals on an ambiguity-controlled test set across different context types and sampling budgets, using a Llama-3.1-8B-Instruct model fine-tuned for intent translation on a vendor-specific switch platform (Juniper EX3300). The results demonstrate that predictive uncertainty provides a useful signal for ranking translations by risk across context types and sampling budgets, albeit with substantial miscalibration under less informative contexts. Moreover, we show that parameter-token entropy correlates with parameter-sourced ambiguity and keyword-token entropy correlates with description-sourced ambiguity.  
These results indicate the potential of using uncertainty signals in an LLM-generated configuration deployment pipeline, where predictive uncertainty can support selective deployment, while token-level entropy can identify sources of ambiguity.  
\end{abstract}

\section{Introduction}
Intent-based networking (IBN) aims to simplify network management by enabling operators to describe networking tasks in high-level intents, while the IBN system handles the implementation details. Intent realization starts by translating intents into low-level configuration, followed by intent verification, conflict detection and resolution, and then deployment. After deployment, the IBN system continuously monitors the network for intent drift during the intent assurance stage and adjusts the network state to align with the user’s intent. Together, these stages constitute the intent life-cycle. 

Large language models (LLMs) have advanced capabilities for understanding natural language and demonstrate versatility across a wide range of domains and applications. Many studies have explored integrating LLMs at different stages of the intent life cycle, including intent translation. Several techniques have been explored for intent translation, such as prompt engineering \cite{wang2024netconfeval, lira2024large, fuad2024intent, jeong2024s, angi25}, fine-tuning \cite{tu2025intent, LiraFT26, PreconfLi}, and retrieval-augmented generation (RAG) combined with fine-tuning \cite{wang2026llm, LiVTC,fang2024llmndc}. Generating low-level configuration is challenging because it follows (\textit{i}) predefined syntax and (\textit{ii}) requires valid parameter values. Accurate translation depends on correctly interpreting the user’s intent, and there is rarely a one-to-one mapping between intents and network configurations. Moreover, ambiguity in intent expression makes translation more challenging and can lead to erroneous configurations. Ambiguity can arise from insufficient parameter values, varying technical expertise, or vendor-specific terminology. For instance, Juniper uses the keyword \texttt{persistent-learning} in the configuration required to enable a port security feature, which keeps learned media access control (MAC) addresses persistent even after a reboot or link-down events. This feature reduces the time required to learn interface-MAC bindings and restores the service quickly. Cisco has a comparable feature, called sticky MAC, configured with the keyword \texttt{sticky}. Such discrepancies can introduce ambiguity when a network operator uses cross-vendor terminology in their intent. 

Existing LLM-based intent translation approaches mainly focus on improving configuration generation accuracy and rely on downstream validation, such as Batfish \cite{mondal2023llms, lira2024large, PreconfLi}. However, less attention has been paid to assessing whether a translation is risky before it reaches verification and deployment. Quantifying LLM uncertainty can estimate completion reliability \cite{liu2025uncertainty}, and several approaches have been proposed for LLM uncertainty quantification, including token-level, self-verbalization, semantic entropy \cite{kuhn2023semantic}, and mechanistic interpretability \cite{shorinwa2025survey}. 
This work investigates whether model uncertainty can serve as a pre-deployment risk signal for LLM-based intent translation. We ask two complementary questions, i.e. which translations are risky, and where within a configuration the ambiguity originates. We study sampling-based predictive uncertainty for ranking translations by risk, and token-level entropy for localizing uncertainty associated with parameters and description ambiguity. Using our fine-tuned Llama-3.1-8B-Instruct model, we evaluate both uncertainty signals on an ambiguity-controlled test set under four context types and three sampling budgets. We address two research questions. \textit{RQ1}: Does predictive uncertainty provide a useful pre-deployment signal for ranking LLM-generated configurations by translation risk? \textit{RQ2}: Can token-level entropy localize ambiguity sources within a generated configuration? 

Our main contributions are as follows.
\begin{itemize}
    \item  We show that sampling-based predictive uncertainty preserves a useful ranking of translation risk across context types and sampling budgets, enabling higher-risk configurations to be prioritized for abstention despite substantial miscalibration.
    \item  We show that token-level entropy separates ambiguity sources, where parameter-token entropy correlates with parameter-sourced ambiguity, and keyword-token entropy correlates with description-sourced ambiguity.
    \item We evaluate both uncertainty signals across sampling budgets of 5, 10, and 30 generated configurations, showing that smaller sampling budgets reduce sampling cost while largely preserving translation accuracy, risk-ranking performance, and ambiguity localization signals.
\end{itemize}
\section{Related Work}
Existing intent translation approaches assess configuration reliability via syntax verification \cite{lira2024large}, simulation-based verification \cite{Intent-LLM}, and human review \cite{tu2025intent}. However, syntax verification cannot determine whether a valid configuration satisfies the user’s intent, simulation requires executing the configuration in a modeled environment, and manual review is labor-intensive and difficult to scale. These limitations motivate the investigation of an inference-time signal to identify risky translations before downstream verification and deployment. 

Ambiguous or underspecified intents can lead to configurations that do not satisfy the user’s intent. Jacobs et al. formally define ambiguity as a pairwise directional relationship between a new intent and deployed intents when no network state can satisfy both \cite{jacobs2025establishing}. They train a random forest classifier on hand-labeled intent pairs to classify them as ambiguous or ambiguity-free, leaving resolution to the operator. Similarly, Mondal et al. define ambiguity as the underspecification of how to handle the insertion of a new access control or routing rule when it overlaps with existing rules \cite{mondal2025tackling}. Their approach generates the configuration for the new intent in isolation, then verifies it using Batfish. Other studies mitigate ambiguity during intent translation. Liu et al. \cite{Liu2025} use chain-of-thought (CoT) prompting \cite{wei2022chain} to analyze the user’s intent, extract key information such as protocols and parameters, decompose complex problems, and rewrite the ambiguous intent into a structured configuration intent. Angi et al. address intent ambiguity implicitly using CoT prompting, few-shot learning, semantic-similarity-based prompt routing, and low-temperature sampling \cite{angi25}. Left open, however, is whether intrinsic model uncertainty can identify ambiguity within a single intent and distinguish parameter-sourced from description-sourced ambiguity.

Uncertainty quantification methods estimate the reliability of LLM completions. Sampling-based methods generate multiple completions and measure their consistency, whereas token-level entropy yields a localized estimate from the probability distribution over generated tokens \cite{liu2025uncertainty}. Neither has been applied to network configuration, where the relevant questions are whether uncertainty can rank generated configurations by translation risk and whether it can localize the source of ambiguity. We evaluate sampling-based predictive uncertainty for the former and token-level entropy for the latter.

\section{Uncertainty Signals for Intent Translation}
Deploying LLM-generated configurations without human review requires identifying likely-incorrect outputs before deployment. Correctness, however, is unavailable at inference time, and syntax verifiers catch malformed configurations but cannot establish whether a well-formed one satisfies the intended task. We therefore formulate uncertainty assessment for intent translation as two complementary problems: translation risk ranking and ambiguity-source localization.

Given a natural-language intent $x$ and $K$ sampled configurations
$\{\hat{y}_1, \hat{y}_2, \dots, \hat{y}_K\}$, where $K$ is the sampling
budget, we investigate two entropy-based signals. The first,
sampling-based predictive uncertainty, quantifies variation across the
$K$ samples and serves as the risk-ranking signal. The second,
token-level entropy, flags individual words within a generated
configuration and is evaluated for localizing the source of
ambiguity as either parameter-sourced or description-sourced. We assess
both signals across context types, as defined in
Section~\ref{context}, and sampling budgets to test their robustness.

\subsection{Sampling-Based Predictive Uncertainty} \label{PU}
Predictive uncertainty characterizes the uncertainty associated with the prediction $\hat{y}$ for query $x$ \cite{hullermeier2021aleatoric}. We measure predictive uncertainty as the normalized Shannon entropy of the empirical distribution over sampled completions, grouping
completions by exact string match. Although semantic clustering can capture equivalent outputs \cite{kuhn2023semantic}, we use string equality because NIT provides only one reference configuration per intent.
Let $\mathcal{C}$ denote the set of distinct configurations among the $K$ samples, and let $p(c)= \frac{n_{c}}{K}$, where $n_{c}$ is the number of times configuration $c$ occurs among the samples. Predictive uncertainty over the empirical completion distribution is thus

\begin{equation}
U(x) = \frac{-\sum_{c \in \mathcal{C}} p(c) \log p(c)}{\log K},
\end{equation} where $U(x) \in [0,1]$. When all $K$ samples produce the same configuration, $U(x) = 0$, whereas $U(x) = 1$ when all sampled configurations are distinct. We define the confidence score derived from predictive uncertainty as 
\begin{equation}
s(x) = 1 - U(x).
\end{equation}

To evaluate predictive uncertainty as a translation risk-ranking signal, let us first evaluate each sampled configuration $\hat{y}_k$ against a reference configuration $y^*$ using an indicator, such that
\begin{equation} 
\mathbf{1}(\hat{y}_k, y^*) = \begin{cases} 1, & \text{if } \hat{y}_k = y^*,\\ 0, & \text{otherwise.} \end{cases} 
\end{equation}
For each intent $x$, configuration generation accuracy is the mean indicator value across the $K$ sampled completions, 

\begin{equation} 
\label{eq:em}
\mathrm{EM}(x) = \frac{1}{K} \sum_{k=1}^{K} \mathbf{1}(\hat{y}_k, y^*). 
\end{equation}
The corresponding translation risk is
\begin{equation} 
R(x) = 1 - \mathrm{EM}(x). 
\end{equation}

\subsection{Token-Level Entropy} \label{entropy-class}
Sampling-based predictive uncertainty captures output-level consistency
but does not reveal which parts of a configuration the model is
uncertain about. We therefore investigate whether token-level entropy
can localize uncertainty to the source of ambiguity that produced it.
For a generated configuration $\hat{y}$, the Shannon entropy at position
$t$ over the model's vocabulary $\mathcal{V}$ is
\begin{equation}
\label{shannon}
H(t) = -\sum_{v \in \mathcal{V}}
p(v \mid x, \hat{y}_{<t})
\log p(v \mid x, \hat{y}_{<t}),
\end{equation}
where $p(v \mid x, \hat{y}_{<t})$ is the probability of token $v$ at
position $t$ conditioned on the intent $x$ and the previously generated
tokens $\hat{y}_{<t}$.

Our experiments use Llama-3.1-8B-Instruct fine-tuned on the NIT dataset
for Juniper EX3300 intent translation \cite{ala2025nit}, which employs a
Byte-Pair Encoding (BPE) tokenizer. BPE splits words into subword units, so an interface name such as \texttt{ge-0/0/10} may span several tokens. Subword entropy therefore
does not directly support the word-level interpretation our ambiguity
analysis requires. We therefore aggregate the subword tokens $\mathcal{T}(w)$ constituting each configuration word
$w$ as
\begin{equation}
H(w) = \max_{t \in \mathcal{T}(w)} H(t),
\end{equation}
taking the most uncertain constituent token as the entropy of the word.

We then classify configuration words as keywords or parameters using a
predefined set of Juniper EX3300 CLI keywords. A keyword belongs to the
vendor's fixed syntax, whereas a parameter is a user-specified value.
Separating the two allows uncertainty to be attributed to distinct
sources: high entropy on parameters suggests underspecified values in
the intent, while high entropy on keywords suggests uncertainty in the
task description or the vendor syntax itself.

Let $\mathcal{K}_k$ and $\mathcal{P}_k$ denote the keyword and parameter
words in sample $\hat{y}_k$. We average word-level entropy within each
category per sample, then across the $K$ samples drawn for intent $x$,
giving the keyword-token entropy
\begin{equation}
H_{\mathcal{K}}(x) = \frac{1}{K} \sum_{k=1}^{K}
\frac{1}{|\mathcal{K}_k|} \sum_{w \in \mathcal{K}_k} H(w),
\end{equation}
and the parameter-token entropy $H_{\mathcal{P}}(x)$, defined
analogously over $\mathcal{P}_k$. Both are computed from the same $K$ samples used for predictive uncertainty in Subsection~\ref{PU}.
Commands that take no variable parameters have $|\mathcal{P}_k| = 0$, rendering $H_{\mathcal{P}}(x)$ undefined. Thus, we exclude the corresponding intents from the parameter-token entropy analysis.

\subsection{Ambiguity Scoring Rubric}
Ambiguity in network intents arises from underspecified parameters or task descriptions. To analyze how different levels of ambiguity affect model uncertainty and translation quality, we define a six-level ambiguity-scoring rubric (L0-L5) as shown in Table \ref{tab:annotation_rubric}. The total ambiguity level is computed as $L = A_P + A_D$, where $A_P$ is the parameter ambiguity score and $A_D$ is the description ambiguity score. A score of 0 indicates that the intent is clear, all required parameters are provided, and it corresponds to the primary functionality of the low-level command. In contrast, a score of 5 indicates highly ambiguous intent unrelated to the command's well-known functionality, with none of the required parameters provided. 

We use Claude Sonnet 4.6 with the ambiguity-scoring rubric to annotate the question field in the NIT dataset entries. We manually reviewed all annotations for consistency with the rubric. The NIT dataset is skewed toward low ambiguity levels and is distributed as follows: 72.1\% of samples at level 0, 22.2\% at level 1, 4\% at level 2, 1.7\% at levels 3 and 4 combined, and no samples at level 5.


\begin{table*}[t]
\vspace*{2.5pt}
\centering
\caption{Annotation Rubric: Parameter and Description Scores.}
\label{tab:annotation_rubric}

\begin{tabularx}{\textwidth}{
    @{}
    l
    >{\raggedright\arraybackslash}X
    c
    @{}
}
\toprule
\textbf{Score Type} & \textbf{Description} & \textbf{Score} \\
\midrule

\multirow{3}{*}{\textbf{Parameter Score ($A_P$)}}
& All required parameters are provided, or the command requires no variable parameters & 0 \\
& At least one parameter value is provided, and at least one is missing & 1 \\
& No parameter values are provided & 2 \\
\midrule

\multirow{4}{*}{\textbf{Description Score ($A_D$)}}
& The task directly maps to the well-known/primary use case of the command in the answer & 0 \\
& Not $A_D=0$, and the expression contains $\geq 2$ keyword mappings & 1 \\
& Not $A_D=0$, and the expression contains exactly 1 keyword mapping & 2 \\
& Not $A_D=0$, and the question is unrelated to the well-known functionality and does not meet the keyword thresholds & 3 \\
\bottomrule
\end{tabularx}
\end{table*}

\subsection{Ambiguity-Controlled Test Split Generation}
To investigate the correlation between sources of ambiguity and token-level entropy, and how disambiguating context affects translation quality, we generate an ambiguity-controlled test split.  We retain 59 entries from the NIT dataset and synthesize an additional 139 entries, as most entries in the dataset are skewed toward clear intents. To ensure alignment between the ambiguity-controlled test split and the NIT design, we follow the same rules as reported in \cite{ala2025nit}. Parameter placeholders were inserted into the referenced command where the intent omitted necessary values. The resulting entries were checked against Juniper EX3300 documentation for syntactic correctness and reviewed by a domain expert for intent satisfaction. 

To introduce parameter ambiguity in the intent, we either omit parameter values or use indirect references, since no topology context is provided during inference. For example, the intent \textit{I need to watch real-time traffic
statistics on the port serving our core switch uplink} is
underspecified, since the intent indirectly references the port but does not specify a value required for translation. On the other hand, description ambiguity is introduced by manipulating keywords and task descriptions while still providing adequate domain context to map the intent to the correct command, even at higher levels of ambiguity. The resulting test split has 46 entries at L0, 31 entries at each of L1 and L2, and 30 entries at each of the remaining ambiguity levels. This enables a systematic analysis of model performance across a wider range of ambiguity than is present in the original NIT dataset. In summary, the 941 entries from the NIT were used for fine-tuning, while the new test split was used only for testing. 

\subsection{Context Types} \label{context}
To analyze how disambiguating context affects uncertainty signals, we
define four context types of increasing informativeness. In the
\emph{none} setting, the model receives only the system prompt and must
infer the entire command from the intent. The \emph{keywords} context
supplies the top two levels of command keywords from the low-level
command, i.e., the command category and topic, which constrain the
command family without specifying values. The \emph{parameters} context
instead supplies the required values as name--value pairs, leaving the
command structure to be inferred. These two are complementary: the
keywords context targets description-sourced ambiguity, whereas the
parameters context targets parameter-sourced ambiguity. Finally, the
\emph{template} context provides the generic form of the command
required to achieve the user's intent.

Because the context determines what the model can be expected to
produce, each context type requires its own reference configuration.
Only the parameters context yields a fully resolved command; the
remaining three retain placeholders, since a model given no user name
cannot be penalized under exact match for emitting
\texttt{<user-name>}. We therefore add a separate answer field, per
context type, to both the NIT dataset and the ambiguity-controlled test
split. Table~\ref{tab:contextEX} illustrates the context values and
corresponding reference commands for a single intent.

\begin{table*}[t]
    \centering
    \caption{Example context types and their corresponding reference commands.}
    \label{tab:contextEX}
    \scriptsize
    \begin{tabular}{@{}l l@{}}
        \toprule
        \multicolumn{2}{c}{\textbf{Intent:} Create new user account with operator privilege} \\
        \midrule
        
        \multirow{2}{*}{\textbf{None}}   & \makebox[2.6cm][l]{Context value}     \texttt{Empty} \\
                                              & \makebox[2.6cm][l]{Reference command} \texttt{set system login user <user-name> class operator} \\\midrule
                                              
        \multirow{2}{*}{\textbf{Keywords}}   & \makebox[2.6cm][l]{Context value}     \texttt{\{cmd\_category: set, topic: system\}} \\
                                              & \makebox[2.6cm][l]{Reference command} \texttt{set system login user <user-name> class operator} \\
        \midrule
        \multirow{2}{*}{\textbf{Parameters}} & \makebox[2.6cm][l]{Context value}     \texttt{\{user-name: netops1\}} \\
                                              & \makebox[2.6cm][l]{Reference command} \texttt{set system login user netops1 class operator} \\
        \midrule
        \multirow{2}{*}{\textbf{Template}}   & \makebox[2.6cm][l]{Context value}     \texttt{set system login user <user-name> class operator} \\
                                              & \makebox[2.6cm][l]{Reference command} \texttt{set system login user <user-name> class operator} \\
        \bottomrule
    \end{tabular}
\end{table*}

\subsection{Fine-tuning Setup}
We perform all experiments on Colab Pro+ using an NVIDIA A100-SXM4 GPU with 40 GB of memory. We fine-tune Llama-3.1-8B-Instruct with Low Rank Adaptation (LoRA) on the final subset of the NIT dataset (941 entries), using the hyperparameters shown in Table \ref{tab:finetuning_params}. We fine-tune the model using mixed context types to improve generalization across deployment scenarios with varying contexts and to evaluate the robustness of the uncertainty signals. During each training epoch, we sample from the following distribution: 28.8\% for none context, 13.6\% for parameters context, 28.8\% for keywords context, and 28.8\% for template context. Parameters context has the lowest sampling probability because it is constrained by the number of entries with missing parameter values. We evaluate uncertainty signals across three sample budgets (5, 10, and 30) to assess the trade-off between sampling cost and uncertainty estimation quality, with a sampling temperature of 0.3.

\begin{table}[ht]
\centering
\caption{Fine-Tuning Hyperparameters for Llama~3.1-8B-Instruct.}
\label{tab:finetuning_params}
\resizebox{\columnwidth}{!}{%
\begin{tabular}{p{2cm}ll}
\toprule
\textbf{Category} & \textbf{Parameter} & \textbf{Value} \\
\midrule
\multirow{5}{*}{LoRA} 
  & Rank ($r$)               & 32 \\
  & Alpha ($\alpha$)         & 16 \\
  & Dropout                  & 0.05 \\
  & Target modules           & \texttt{q/k/v/o, gate/up/down\_proj} \\
\midrule
\multirow{7}{*}{Training}
  & Epochs                   & 3 \\
  & Batch size               & 8 \\
  & Learning rate            & $1 \times 10^{-4}$ \\
  & LR scheduler             & Cosine \\
\bottomrule

\end{tabular}}
\end{table}

\subsection{Evaluation Metrics} \label{sec:eval_metr}

Minor deviations from the reference configuration, such as an incorrect
parameter value or a missing command keyword, can cause deployment
failures. Lexical similarity metrics such as ROUGE and BLEU are
therefore insufficient for evaluating configuration correctness.
Consider the following commands:
\begin{center}
\footnotesize
\begin{tabular}{@{}r@{\hspace{5pt}}l@{}}
(a) & \texttt{set ethernet-switching-options secure-access-port} \\
    & \texttt{\quad interface ge-0/0/8 mac-limit \textbf{10}} \\[4pt]
(b) & \texttt{set ethernet-switching-options secure-access-port} \\
    & \texttt{\quad interface ge-0/0/8 mac-limit \textbf{1}} \\
\end{tabular}
\end{center}
They differ in a single character and achieve a ROUGE-1 score of
$0.933$, yet enforce different policies. The first admits up to ten MAC
addresses on the port, while the second admits only one. We therefore use EM to evaluate the correctness of generated configurations.

We evaluate predictive uncertainty from three complementary perspectives. First, we assess calibration using expected calibration error (ECE). A model is well calibrated when its confidence matches its empirical accuracy, i.e., among the intents assigned confidence $s(x) = q$, a fraction $q$ are translated correctly \cite{guo2017calibration}.
Second, we evaluate selective prediction performance using the area under the risk-coverage curve (AURC), which measures how effectively predictive uncertainty ranks translations for uncertainty-based abstention \cite{geifman2017selective}. Finally, we compute Spearman’s rank correlation between predictive uncertainty and EM to assess whether translations with higher uncertainty tend to have lower translation accuracy.

Additionally, we compute Spearman’s rank correlation between token-level entropy and the corresponding ambiguity source score. To isolate the effect of each ambiguity source, we only use entries in which the other ambiguity source is zero. For example, when computing the correlation between parameter-token entropy and parameter ambiguity, we only use entries with a description ambiguity score of zero.

\section{Results and Discussion}
Table~\ref{tab:em} reports the configuration generation accuracy
$\overline{\mathrm{EM}} = \frac{1}{N}\sum_{i=1}^{N}\mathrm{EM}(x_i)$,
where $\mathrm{EM}(\cdot)$ is given in \eqref{eq:em} and $N$ is the number
of intents in the test split, for each context type and sampling
budget.
Reducing $K$ from 30 to 10 and 5 changes the average EM by one percentage point at most, while the sampling cost is reduced by approximately 67\% and 83\%, respectively.

\begin{table}
\vspace*{4pt}
    \centering
    \caption{$\overline{\mathrm{EM}}$ across context types and sampling budgets.}
    \label{tab:em}
    \begin{tabular}{cccc}
    \toprule
     \textbf{ Context Type }  & \textbf{K = 30} & \textbf{K = 10} &\textbf{K = 5} \\ \midrule
      None   &  20\% & 19\%   &20\% \\
      Keywords   & 23\% & 24\% & 24\% \\
      Parameters   & 21\% & 21\% & 21\% \\
      Template   & 67\% & 66\%   & 66\%\\\bottomrule
    \end{tabular}
    
\end{table}

We now report the three perspectives introduced in
Subsection~\ref{sec:eval_metr}. First, Spearman’s correlation between EM and predictive uncertainty is negative across all context types and sampling budgets. It ranges from -0.25 to -0.40 in the none, keywords, and parameters contexts, and is stronger in the template context, ranging from -0.52 to -0.59. Reducing $K$ from 30 to 5 weakens the correlation magnitude by at most 0.09, indicating that the risk-ranking signal is largely preserved at smaller sampling budgets.

\begin{figure*}[!t]
    \centering
    \includegraphics[width=\textwidth]
    {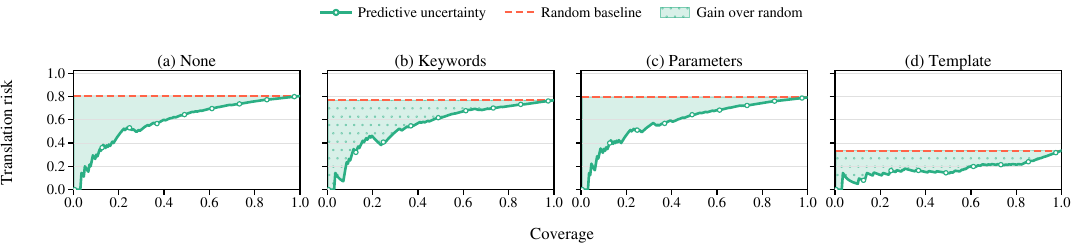}
    \caption{Risk--coverage curves at $K=30$ across context types.}
    \label{fig:risk_coverage}
\end{figure*}

    
Second, we evaluate selective prediction performance by computing the AURC gain of predictive uncertainty relative to an analytical random baseline, represented by a flat risk–coverage curve with constant risk of 
$1 - \overline{\mathrm{EM}}$.
At $K = 30$, predictive uncertainty achieves AURC gains of 26.13\%, 27.25\%, 25.63\%, and 45.21\% under the none, keywords, parameters, and template context types, respectively, as shown in Fig.~\ref{fig:risk_coverage}. Relative to $K=30$, reducing $K$ to 10 or 5 changes the AURC gain by at most 2.88 percentage points, with gains generally decreasing. The only exception is the parameters context at $K = 10$, where the AURC gain increases by 1.27 percentage points.

Lastly, we evaluate the calibration of the confidence score derived from predictive uncertainty using ECE. At $K = 30$, the ECE ranges from 0.56 to 0.59 under the none, keywords, and parameters context types, indicating considerable miscalibration between the model’s confidence and accuracy. Relative to $K = 30$, reducing $K$ to 5 improves calibration and reduces ECE by 0.08 to 0.10. In contrast, the template context yields substantially lower ECE values across all $K$ values, ranging from 0.27 to 0.28. These results indicate that predictive uncertainty is more suitable for relative translation-risk ranking than for estimating absolute confidence.

    

We now turn to token-level entropy, which localizes uncertainty within a configuration. We test whether parameter-token entropy tracks parameter-sourced ambiguity and keyword-token entropy tracks description-sourced ambiguity, and whether these relationships persist across context types and sampling budgets. Tables \ref{tab:parmCorr} and \ref{tab:KeywordCorr} show the correlation between ambiguity source and the corresponding token class. 
Parameter-token entropy correlates strongly with parameter-sourced ambiguity under the template context, at 0.85 for every sampling budget, and moderately elsewhere, from 0.43 to 0.54. Relative to $K = 30$, reducing $K$ to 5 increases the correlation under none and keywords context types, while it remains unchanged under parameters and template context types. Keyword-token entropy correlates weakly-to-moderately with description-sourced ambiguity, from 0.36 to 0.54.

\begin{table}
\vspace*{4pt}
    \centering
    \caption{Spearman’s correlation between parameter-token entropy and parameter-sourced ambiguity across context types and sampling budgets.}
    \label{tab:parmCorr}
    \begin{tabular}{cccc}
    \toprule
     \textbf{ Context Type }  & \textbf{K = 30} & \textbf{K = 10} &\textbf{K = 5} \\ \midrule
      None &   0.43&0.46 &0.50      \\
      Keywords   & 0.47 & 0.46  &0.54  \\
      Parameters   & 0.44& 0.48 &0.44 \\
      Template   &  0.85&0.85   &0.85 \\\bottomrule
    \end{tabular}
    
\end{table}

\begin{table}
    \centering
     \caption{Spearman’s correlation between keyword-token entropy and description-sourced ambiguity across context types and sampling budgets.}
    \label{tab:KeywordCorr}
    \begin{tabular}{cccc}
    \toprule
     \textbf{ Context Type }  & \textbf{K = 30} & \textbf{K = 10} &\textbf{K = 5} \\ \midrule
      None &   0.53& 0.54 &0.52     \\
      Keywords  &0.36&0.39&0.36    \\
      Parameters & 0.53&0.54&0.52  \\
      Template   & 0.44&0.44&0.45 \\\bottomrule
    \end{tabular}
   
\end{table}
In summary, both research questions are answered affirmatively.
Predictive uncertainty ranks configurations by translation risk, as
evidenced by its negative correlation with EM and positive AURC gains, though its high ECE under the none, keywords, and parameters contexts restricts it to relative ranking rather than a proxy for correctness. Token-level entropy correlates positively with its matching ambiguity source in every cell, and smaller sampling budgets preserve both signals at substantially lower cost.
\section{Conclusion}
We investigated whether model uncertainty can serve as a pre-deployment risk signal for LLM-based intent translation, where we proposed sampling-based predictive uncertainty for translation-risk ranking and token-level entropy for ambiguity-source localization. The two signals suggest a concrete role in a deployment pipeline. Predictive uncertainty ranks reliably but is poorly calibrated in our setting, since a confidence score derived from sample agreement cannot detect errors the model makes consistently. It can therefore be used to select \emph{which} configurations to withhold for review, ranking them so that an operator can tune the abstention rate against available review capacity. Establishing an absolute confidence threshold for an automated go/no-go deployment decision requires further work, as it depends on calibration quality that the present estimator does not provide. Token-level entropy can be used to indicate what to do with a withheld configuration, as entropy concentrated on parameters points to a missing value the operator can supply, whereas entropy on keywords points to a misread intent that requires rephrasing. This distinction is what would make clarification actionable rather than a generic request to restate the intent.

Future work can consider implementing the uncertainty-aware deployment pipeline these results motivate, coupling abstention thresholds to entropy-driven clarification generation. Future work can also consider examining confidence estimators that account for semantic equivalence rather than string agreement alone, and to test whether both signals hold across vendors and model scales.

\bibliographystyle{IEEEtran}
\bibliography{ref}

\vfill

\end{document}